# Pore-Scale Visualization of Hydrogen Storage in a Sandstone at Subsurface Pressure and Temperature Conditions: Trapping, Dissolution and Wettability

Zaid Jangda[1,*], Hannah Menke[1], Andreas Busch[1], Sebastian Geiger[2], Tom Bultreys[3], Helen Lewis[1], Kamaljit Singh[1]

[1]Institute of GeoEnergy Engineering, Heriot-Watt University, EH14 4AS, Edinburgh, United Kingdom
[2]Department of Geoscience and Engineering, Delft University of Technology, 2628 CN Delft, Netherlands
[3]UGCT/PProGRess, Department of Geology, Ghent University, 9000 Ghent, Belgium

*Corresponding author

Email addresses: zj21@hw.ac.uk (Jangda. Z), h.menke@hw.ac.uk (H. Menke), a.busch@hw.ac.uk (Busch. A), s.geiger@tudelft.nl (Geiger. S), tom.bultreys@ugent.be (Bultreys. T), h.lewis@hw.ac.uk (Lewis. H), k.singh@hw.ac.uk (Singh. K)

Phone: +44 7761955975 (Jangda. Z)

## Graphical Abstract

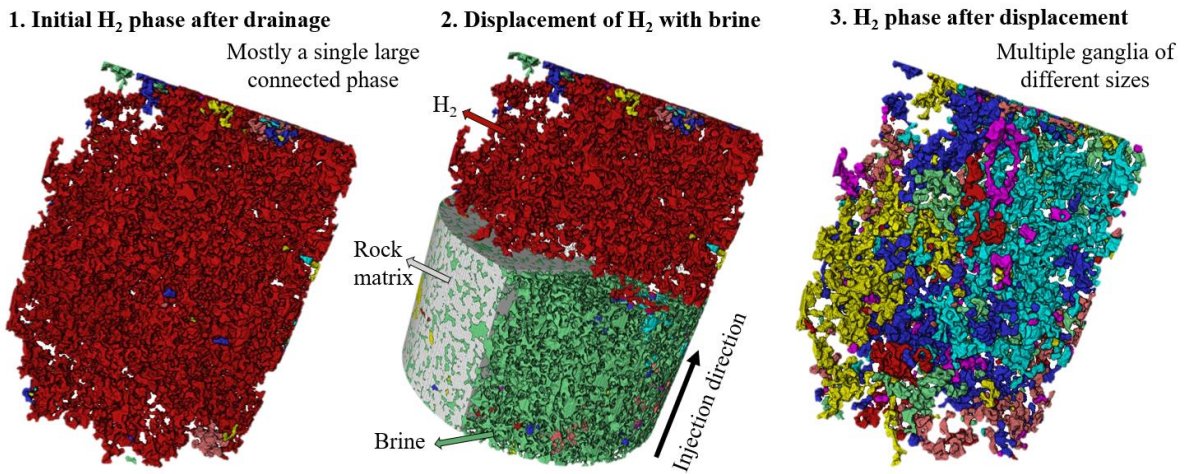

1. Initial $H_2$ phase after drainage — Mostly a single large connected phase
2. Displacement of $H_2$ with brine — $H_2$, Rock matrix, Brine, Injection direction
3. $H_2$ phase after displacement — Multiple ganglia of different sizes

## Abstract

### *Hypothesis*

The global commitment to achieve net-zero has led to increasing investment towards the production and usage of green hydrogen ($H_2$). However, the massive quantity needed to match future demand will require new storage facilities. Underground storage of $H_2$ is a potentially viable solution, but poses unique challenges due to $H_2$'s distinctive physical and chemical properties, that have yet to be studied quantitatively in the subsurface environment.

### *Experiments*

We have performed *in situ* X-ray flow experiments to investigate the fundamentals of pore-scale fluid displacement processes during $H_2$ injection into an initially brine saturated Bentheimer sandstone sample. Two different injection schemes were followed, the displacement of $H_2$ with $H_2$-equilibrated brine and non-$H_2$-equilibrated brine both at temperature and pressure conditions representative of deep underground reservoirs.

*Findings*

$H_2$ was found to be non-wetting to brine after both displacement cycles, with average contact angles between 53.72° and 52.72°, respectively. We also found a higher recovery of $H_2$ (43.1%) for non-$H_2$-equilibrated brine compared to that of $H_2$-equilibrated brine (31.6%), indicating potential dissolution of $H_2$ in unequilibrated brine at reservoir conditions. Our results suggest that $H_2$ storage may indeed be a suitable strategy for energy storage, but considerable further research is needed to fully comprehend the pore-scale interactions at reservoir conditions.

# 1. Introduction

The Intergovernmental Panel on Climate Change (IPCC)[1] has warned that the world is set to reach the 1.5°C warming level within the next two decades, and that only the most drastic and immediate cuts in carbon emissions will prevent an environmental disaster. Clean energy alternatives such as $H_2$ have the potential to support near-zero greenhouse gas emissions[2–7]. While there has been an unprecedented economic and political momentum towards the use of green $H_2$, the massive amount needed for future demand would require enormous storage facilities[8–12]. An underground $H_2$ storage capacity between 250-1000 TWh is forecasted to be required for Europe by 2050[13].

Underground $H_2$ storage (UHS), e.g., in porous rocks or salt caverns[13–16], can provide an effective solution for short- and long-term storage to meet the fluctuations in energy demand. Salt caverns have already been used for storing natural gas and $H_2$, e.g., at Cheshire, Teesside, and Yorkshire[17] in the UK; however, they have a limited storage capacity[14]. To scale up for future demand, geological formations, such as depleted gas reservoirs[13,18,19] and saline aquifers that are ubiquitous, have the potential to store large volumes of $H_2$, the feasibility of which is currently being investigated in many countries[10,12,13,17,20].

These underground storage facilities are regarded as economically viable with substantially less capital cost and a lower levelized cost of $H_2$ storage (1.23-1.26 $/kg) compared to that of salt caverns (1.61 $/kg)[21]. However, there are significant uncertainties in storing $H_2$ in rocks[22,23], which is due to the limited available knowledge to use for predicting (i) the uncontrolled migration and unwanted trapping of $H_2$ in reservoir rocks[23–25], (ii) reactivity with water and minerals[26–28], (iii) diffusion through the reservoir and caprock[8,29], and (iv) formation of biofilms[30,31] and their impacts on pore blockage and storage efficiency. Capillary trapping and wettability are of particular interest for $H_2$ as the amount of trapping directly affects recovery efficiency[32–34].

Capillary trapping mechanisms have been extensively studied for $CO_2$, $N_2$, and oil in various porous media[35–42]. These fundamental studies are generally focused on the pore-scale, where the capillary trapping can be observed and quantified directly. However, it is not yet known if $H_2$ behaves similarly in subsurface rocks, where capillary, viscous, and buoyancy forces are likely to play a key role along with other geochemical and biological processes. Most of the available studies on $H_2$ storage in rocks are focused on reservoir-scale numerical simulations[16,43–46], which do not provide pore-scale information.

In addition, there is still a lack of consensus on the wettability state for subsurface $H_2$ storage systems. A few recent experimental studies have reported a wide range of contact angle values for $H_2$-brine systems. Using the tilted plate technique, contact angles of 0°-50° for pure quartz and 50°-70° for a quartz sample aged with stearic acid were measured for a pressure range of 1-250 bar and a temperature range of 20-70°C[34]. A clear trend of increasing contact angle with an increase in pressure and temperature was observed. It was also reported[26] that aging in stearic acid increased the contact angles for $H_2$-brine system from strongly water-wet to intermediate-wet for calcite, dolomite, and shale and from strongly to weakly water-wet for gypsum, quartz, and basalt. These measurements were performed using the captive-bubble method between temperatures of 20-80°C and pressures of 10-100 bar. These studies conclude that under subsurface conditions, $H_2$ wettability is prone to be intermediate-wet for reservoir rocks.

In contrast, contact angle values between 21.1° and 43° have been reported for experiments performed on Berea and Bentheimer sandstone samples using the captive bubble method at a temperature range of 20-50°C and a pressure range of 20-100 bar[24]. The contact angle was not affected by the changes in temperature, pressure, salinity, and rock type; however, it was affected by the bubble size. Similarly, advancing contact angles ranging between 13°-39° and receding contact angles between 6°-23° have been reported for a $H_2$-water system using a microfluidic device with varying channel widths at a temperature of 20°C and pressure of 10 bar[47]. Moreover, contact angles of 21.56° and 34.9° at two different experimental conditions of 50 bar, 20°C and 100 bar, 45°C respectively were reported based on core flood experiments in Voges sandstone[48]. These studies indicate strong water-wet conditions under all the tested parameters. However, to date no study has measured *in situ* contact angles of $H_2$-brine systems at reservoir conditions in a real rock.

*In situ* measurements have been made possible by using X-ray micro-tomography to visualize multiphase fluid displacement events at the pore-scale within porous media [40,49–53]. The images and measurements obtained using this technique have contributed enormously towards developing more efficient strategies for hydrocarbon recoveries and underground $CO_2$ storage[35,39]. A few recent studies have used this technology to investigate the pore-scale behaviour[32,54] (wettability and trapping) of $H_2$; however, these studies have been conducted at ambient conditions which do not represent the actual subsurface storage conditions.

In this study, we have performed $H_2$ core-flood experiments in a sandstone sample at temperature and pressure conditions of 50°C and 100 bar respectively using a custom-designed flow apparatus and imaged the core *in situ* in an X-ray micro-CT scanner. The sample was imaged before and after drainage and imbibition, followed by segmentation and image analysis to quantify the phase volumes, $H_2$ saturation, trapping, and contact angles. We provide a quantitative analysis of the amount of $H_2$ trapped in the subsurface during underground $H_2$ storage and determine contact angle *in situ*. Furthermore, we examine the potential effect of dissolution of $H_2$ by providing a comparison of the $H_2$ residual saturations when using $H_2$ equilibrated brine and non-$H_2$ equilibrated brine as the imbibing fluid.

## 2. Materials and Methods

*2.1 Equipment and Materials*

A 6 mm diameter and 27 mm long water-wet Bentheimer sandstone sample (typical composition of Quartz: 91.7%, Feldspar: 4.86%, Clay: 2.68%, Pyrite and Iron Hydroxides: 0.17%)[55] was used as the porous medium. The sample was cleaned with methanol and dried in a vacuum oven at 100°C for 18 hours. A solution of 4 wt.% potassium iodide (KI) salt (Sigma–Aldrich, UK) in deionized water was used as the aqueous phase, which provided an effective X-ray contrast between brine (wetting phase) and $H_2$ (non-wetting phase). High purity (>99.99%) $H_2$ gas (BOC) was used as the gas (non-wetting) phase.

The experiments were performed using a custom-designed flow apparatus as shown in Figure 1. It consists of a high-pressure and high-temperature core holder (rs systems), high-pressure pumps (ISCO, three model 500D, and one model 100DX), and a reactor (Parr Instruments Co., IL, USA). The sample was wrapped with Teflon tape and aluminium foil and placed inside a Viton sleeve connected to the flow lines. The PEEK core holder was wrapped with a heating tape and securely fitted on top of the rotation stage of the X-ray micro-CT scanner (EasyTom 150, RX Solutions).

A thermocouple was inserted through one of the ports from the base of the core holder and attached outside the Viton sleeve with aluminium tape to measure the temperature outside the rock sample. To control the overtemperature of the system, another thermocouple was attached to the heating tape wrapped around the core holder. Four high-precision syringe pumps were used to apply confining pressure, and inject and receive the fluids. A Hastelloy reactor was included in the flow loop to pre-equilibrate the brine with $H_2$ at the experimental pressure and temperature conditions to avoid any possible dissolution of $H_2$ in the brine during injection in cycle 1 (discussed in the forthcoming section).

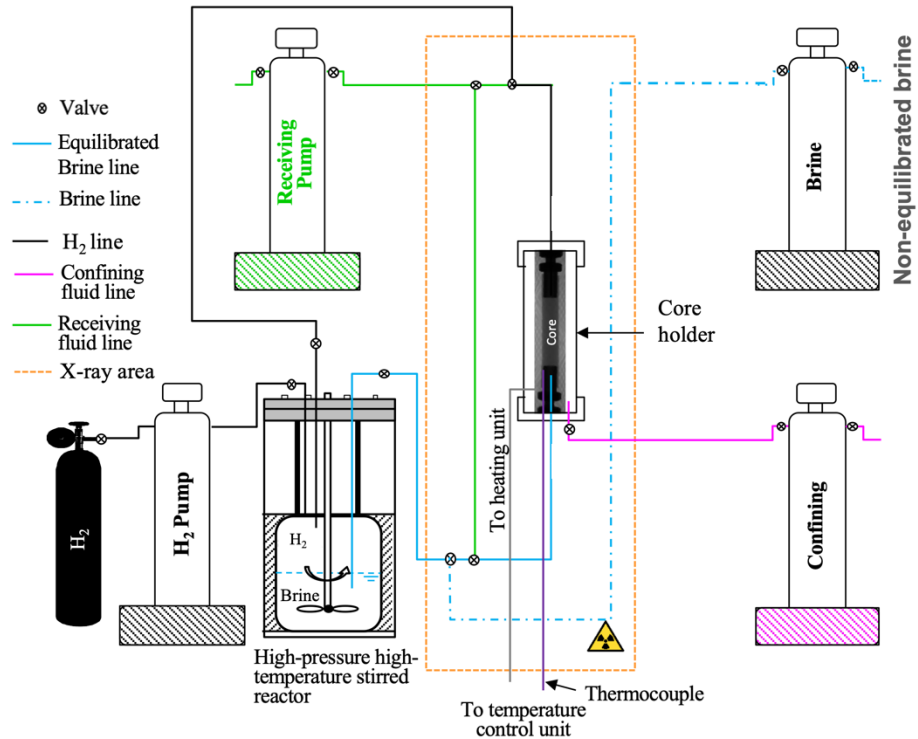

**Figure 1. Experimental apparatus.** A sketch of the experimental flow apparatus showing different components.

## 2.2 Experimental Methodology

The experimental protocol consisted of the following steps:

i. loaded the sample into the core holder
ii. applied a confining pressure of 20 bar
iii. flushed the sample with $CO_2$ from the base to displace any air inside the sample, followed by 75 pore volumes (PV) ($1\ PV = 0.19$ mL) of non-$H_2$-equilibrated brine injection from the base at 0.5 mL/min to remove $CO_2$ dissolved brine, ensuring 100% brine saturation.
iv. switched on the heating to gradually reach 50°C.
v. pressurised the system with non-$H_2$-equilibrated brine to 100 bar in steps, while maintaining a 20-bar difference between the pore and the confining fluid.
vi. injected 70 PV of $H_2$-equilibrated brine from the reactor to completely displace non-$H_2$-equilibrated brine from the sample.
vii. cycle1: injected 15 PV of $H_2$ (non-wetting phase) from the reactor to start drainage at a flow rate of 0.05 mL/min ensuring a capillary number (*Ca*) of $4.16 \times 10^{-9}$. In this step, $H_2$ was injected from the top of the sample.
viii. injected 5 PV of $H_2$-equilibrated brine (wetting phase) from the reactor to displace $H_2$ at the same flow rate ensuring a *Ca* of $2.33 \times 10^{-6}$. In this step, brine was injected from the base of the sample.
ix. cycle 2: repeated step vii, i.e., injected 15 PV of $H_2$ from the top of the sample to start the second drainage cycle.

x. injected 5 PV of non-$H_2$-equilibrated brine (wetting phase) from the base of the sample using a pump.

The sample (middle vertical section of 6.7 mm) was scanned after step (ii) to image the dry sample, and after steps (vii)-(x) to visualize the fluid positions after each displacement process. All scans were acquired using X-ray micro-CT with a voxel size of 5 µm, 1792 projections, 90 kV tube voltage and 10 W tube power.

## 2.3 Image Processing

The obtained images were processed and visualized using Avizo 2021.2 (ThermoFisher Scientific) software as shown in Figure 2. A sub-volume of 1250 slices corresponding to a sample length of 6.22 mm was selected for image processing, on which a non-local means filter[56] was applied to remove the noise. All the wet images (with fluids) were registered to the dry reference image using normalized mutual information and then resampled onto the same voxel grid as the dry reference image. Image segmentation was then performed using a watershed algorithm[57] based on the grey-scale intensity of each voxel. The dry sample was segmented into pores and grains which served as the mask for the wet segmented images. The porosity value of 22.8% obtained from the segmented volume is in good agreement with reported Bentheimer porosity values of 20-26%[58,59].

For each of the subsequent images from drainage and imbibition, $H_2$ was segmented using the watershed technique, after which it was subtracted from the pores (from dry sample) to get the brine phase. At some locations, a superficial thin wetting layer between $H_2$ and rock grains was observed due to the partial volume affects associated with low image resolution. The segmentation was refined in multiple steps to ensure removal of this superficial layer, while maintaining the original segmentation (refer to Supplementary Material).

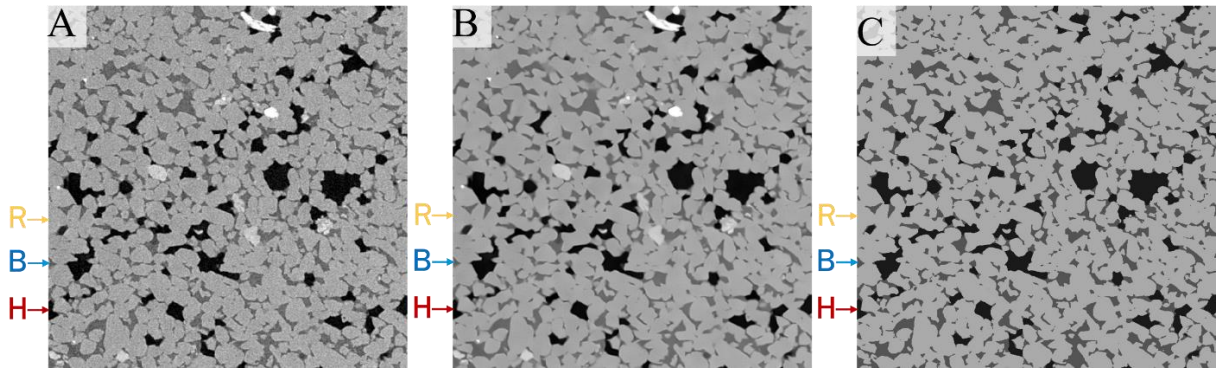

**Figure 2. Image processing workflow.** Raw image (A) was filtered using a non-local means filter (B) for noise removal and then segmented using a seeded Watershed algorithm (C). The segmented phases are $H_2$ 'H' (black), brine 'B' (dark grey) and rock 'R' (light grey).

## 2.4 Saturation, pore occupancy and contact angle

Quantitative analysis was then performed on the segmented images to calculate $H_2$ saturations, pore occupancy and contact angles. $H_2$ phase volume after each displacement process was used to get the initial and residual saturations. To obtain the pore occupancy of $H_2$, first the pores from the dry scan were extracted using a network extraction algorithm[60]. This is based on the modified maximal ball algorithm that extracts simplified networks of pores and throats from the images of the pore space[61], by finding constrictions and wider regions in the pore space, to identify where pores in the image are separated by throat surfaces[62]. These pores (extracted as spheres) were then overlayed on each of the segmented $H_2$ images, to determine the radii and the number of pores filled with $H_2$ after each displacement process.

To ascertain the wettability of the system after each imbibition step, we measured the $H_2$/brine/rock contact angles on each voxel at the three-phase contact line on a sub-volume of 2.5 mm³ using an automated algorithm[63]. This method identifies and meshes the interfaces throughout the segmented image stack and then smooths and adjusts the data to eliminate noise and impose a constant curvature. Two normal vectors are then placed at each contact point and the dot product of these vectors (where they meet at the contact line) is used to determine the contact angle at each contact point along the contact line[63]. This method has been used in multiple studies[64–68] to automatically measure contact angle values for X-ray tomography images with high accuracy.

## 3. Results and Discussion

### 3.1 Drainage and Imbibition cycle 1 ($H_2$-equilibrated brine)

The imaging results of $H_2$ drainage and imbibition with pre-equilibrated brine are shown in Figure 3. The filtered 2D vertical cross-sections in Figure 3A and B show the displacement of $H_2$ with brine, predominantly in the smaller pores after imbibition. Figure 3C shows a well-connected percolating $H_2$ cluster (red) with a few isolated $H_2$ ganglia shown in blue and yellow. During imbibition, $H_2$ fragments into smaller clusters (Figure 3D). The colours in Figure 3C and D represent the different sizes of the $H_2$ ganglia classified based on the average pore size ($P_{avg}$) of $1.12 \times 10^6 \ \mu m^3$, where

- Red – $H_2$ ganglia occupying pore sizes $> P_{avg} \times 10^2$
- Yellow – $H_2$ ganglia occupying pore sizes between $P_{avg} \times 10^1$ and $P_{avg} \times 10^2$
- Blue – $H_2$ ganglia occupying pore sizes $< P_{avg} \times 10^1$

The $H_2$ saturation calculated after drainage was 36.31%, which reduced to 24.83% after imbibition with $H_2$-equilibrated brine, giving a recovery of 31.6%.

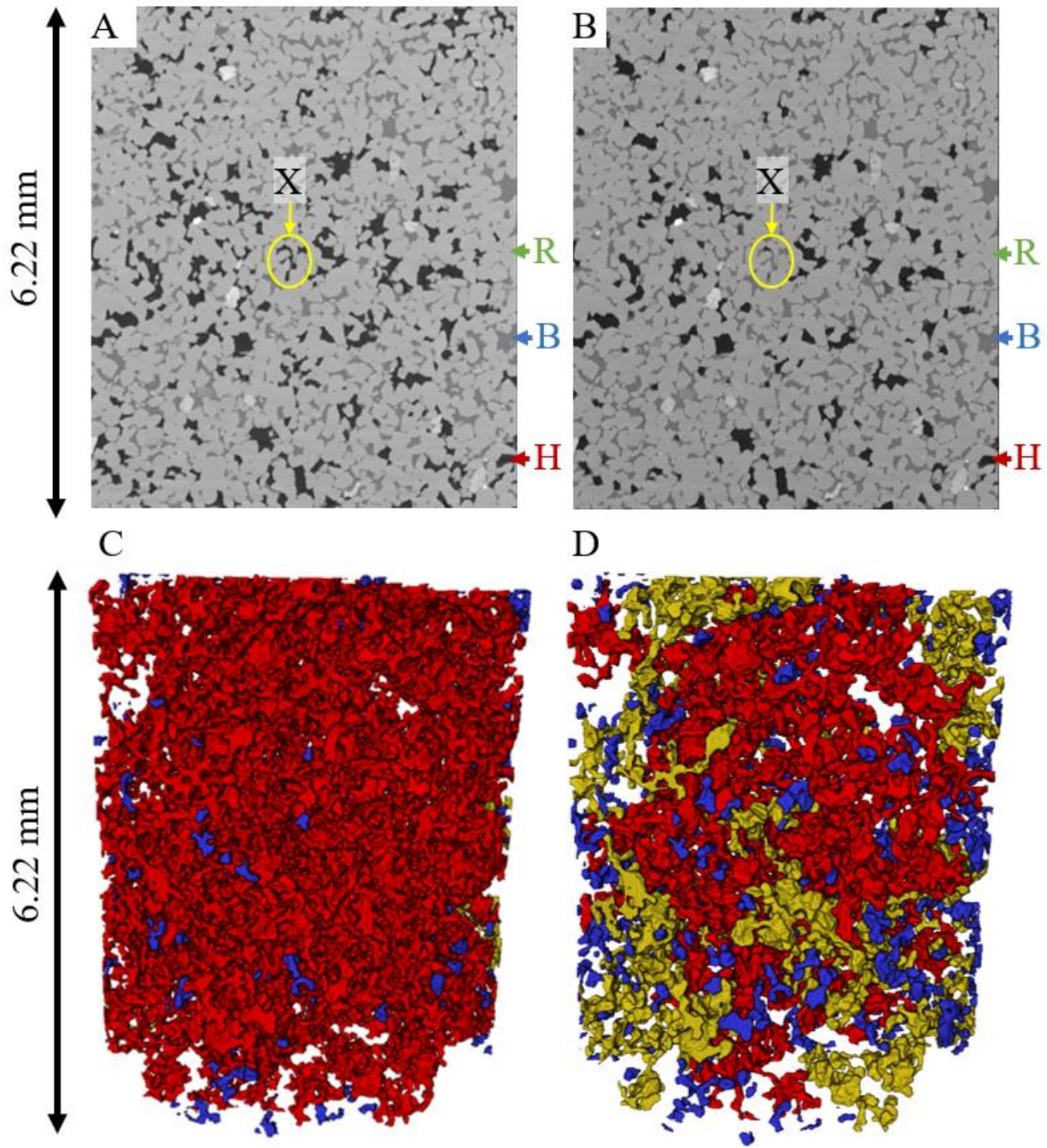

**Figure 3**. **2D and 3D visualization after drainage with H$_2$ and imbibition with H$_2$-equilibrated brine.** (A) 2D vertical cross-section of the filtered image after drainage, (B) 2D vertical cross-section of the filtered image after imbibition, (C) 3D visualization of H$_2$ after drainage and (D) 3D visualization of H$_2$ after imbibition. The 2D images show that after imbibition, the brine 'B' replaces H$_2$ 'H' in the smaller pores (for e.g., the pore space inside the yellow marking represented by 'X'), while H$_2$ remains in the larger pores. 'R' represents the rock grains. The 3D images show that the connected H$_2$ phase after drainage is fragmented into several H$_2$ ganglia after imbibition. Red, yellow, and blue colour respectively represent H$_2$ ganglia occupying pores of sizes that are > 100, between 10-100 and up to 10 times larger than the average pore size.

The $H_2$ pore occupancy is shown in Figure 4. The shift of $H_2$ pore occupancy towards larger pores indicates that the larger pores are occupied by trapped $H_2$ after imbibition. The average pore radius of the pores occupied by $H_2$ increases from 36.27 µm after drainage to 41.14 µm after imbibition.

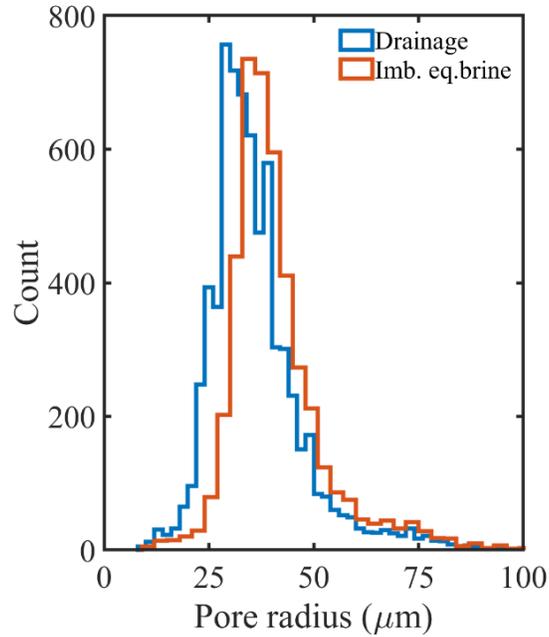

**Figure 4**. **Pore occupancy analysis.** Radius of the pores occupied by $H_2$ after drainage and imbibition with $H_2$-equilibrated brine.

The $H_2$ saturation profile along the height of the sample in the direction of flow is plotted in Figure 5. The saturation varies between 24% and 48% during drainage and 15% and 35% during imbibition, showing some variation in rock heterogeneity in the core. Furthermore, the change in saturation between drainage and imbibition remains constant over the length of the sample indicating the absence of capillary end effects, which is expected as only a 6.2 mm long subsection in the centre of the 27 mm core sample is visualized and analysed.

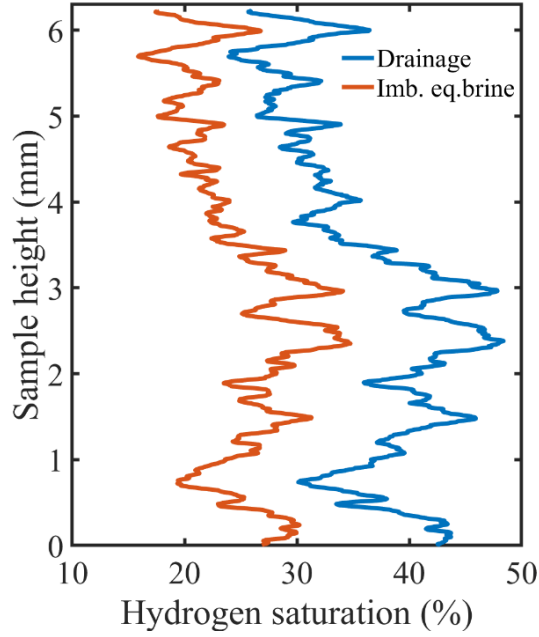

**Figure 5.** Saturation profile along the height of the sample after drainage and imbibition with $H_2$-equilibrated brine.

## 3.2 Drainage and Imbibition cycle 2 (non-$H_2$-equilibrated brine)

The core was then re-saturated with $H_2$-equilibrated brine. In cycle 2, a drainage of 15 PV was performed with $H_2$ and imbibition with 5 PV of non-equilibrated brine (Figure 6). The filtered 2D vertical cross-sections in Figure 6A and B show that brine displaced $H_2$ in a larger number of pores compared to cycle 1. Like cycle 1, $H_2$ mostly exists as a single cluster after drainage (Figure 6C) which is then fragmented into smaller clusters during imbibition (Figure 6D). The colours in Figure 6 represent $H_2$ ganglia occupying pores of different sizes classified based on the average pore size ($P_{avg}$) of $1.12 \times 10^6 \ \mu m^3$

- Red – $H_2$ ganglia occupying pore sizes $> P_{avg} \times 10^2$
- Yellow – $H_2$ ganglia occupying pore sizes between $P_{avg} \times 10^1$ and $P_{avg} \times 10^2$
- Blue – $H_2$ ganglia occupying pore sizes $< P_{avg} \times 10^1$

The $H_2$ saturation after drainage was 35.8%. Residual $H_2$ saturation after imbibition with non-$H_2$-equilibrated brine was measured to be 20.36%, indicating a recovery of 43.1%. However, we posit that it is likely that some of the $H_2$ dissolved in the non-$H_2$-equilibrated brine resulting in lower residual saturation in this cycle. Moreover, isolated clusters of $H_2$ in Figure 6D visually appear different than those in Figure 3D, which will be compared quantitatively in the forthcoming sections.

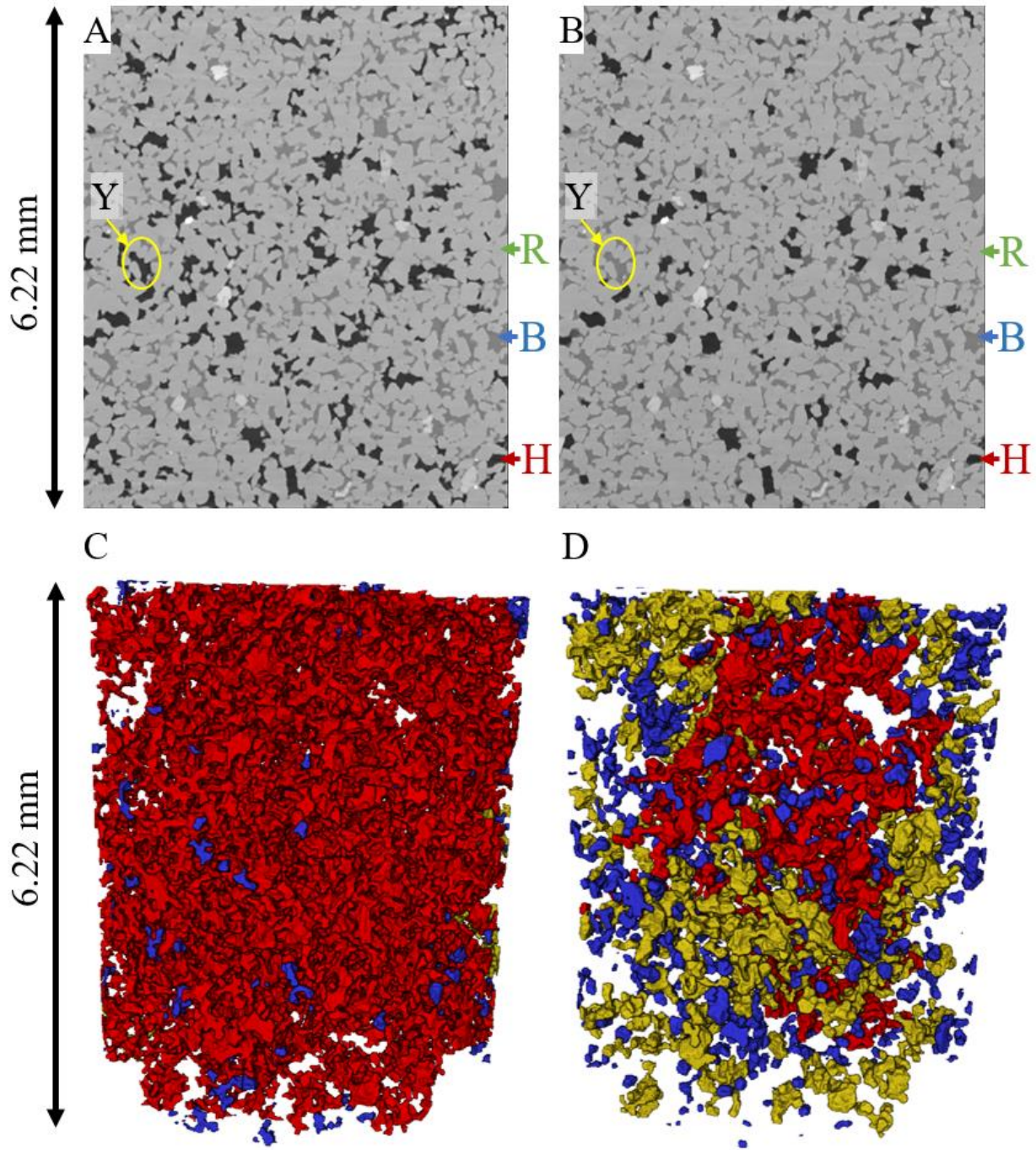

**Figure 6. 2D and 3D visualization after drainage with H$_2$ and imbibition with non-H$_2$-equilibrated brine**. (A) 2D cross-section of the filtered image after drainage, (B) 2D cross-section of the filtered image after imbibition. (C) 3D visualization of H$_2$ after drainage and (D) 3D visualization of H$_2$ after imbibition. The 2D images show that after imbibition, the brine 'B' replaces H$_2$ 'H' even in some of the pores (for e.g., pore space inside the yellow circles represented by 'Y') which remained saturated with H$_2$ after imbibition with H$_2$-equilibrated brine. 'R' represents the rock grains. The 3D images show that the connected H$_2$ phase after drainage is fragmented into several H$_2$ ganglia after imbibition.

Figure 7 shows the pore occupancy of $H_2$ calculated using image analysis for both drainage and imbibition in cycle 2. We observe the same shift as in cycle 1, with larger pores remained occupied by $H_2$ after imbibition. However, the reduction in the $H_2$ pore occupancy after imbibition in cycle 2 is much more enhanced, with $H_2$ occupying a significantly smaller number of pores. The average radius of the pores occupied by $H_2$ (36.14 µm after drainage, and 42.9 µm after imbibition) are similar to the values obtained after cycle 1; however, the reduction in the number of pores occupied by $H_2$ after imbibition is much higher. This shows that although $H_2$ still occupies the larger pores after imbibition, it occupies a significantly lower number of pores, again indicating possible dissolution of $H_2$ in the imbibing brine.

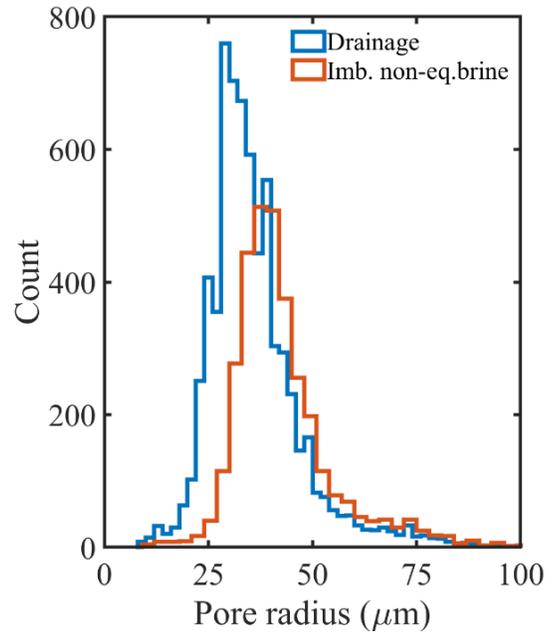

**Figure 7.** Radius of the pores occupied by $H_2$ after drainage and imbibition with non-$H_2$-equilibrated brine.

The $H_2$ saturation profile along the height of the sample is plotted in Figure 8. The saturation varies between 24% and 47% during drainage and 10% and 28% during imbibition. The change in saturation between drainage and imbibition is larger than in cycle 1, indicating a greater loss of $H_2$ from the sample. However, it is still constant over the length of the sample.

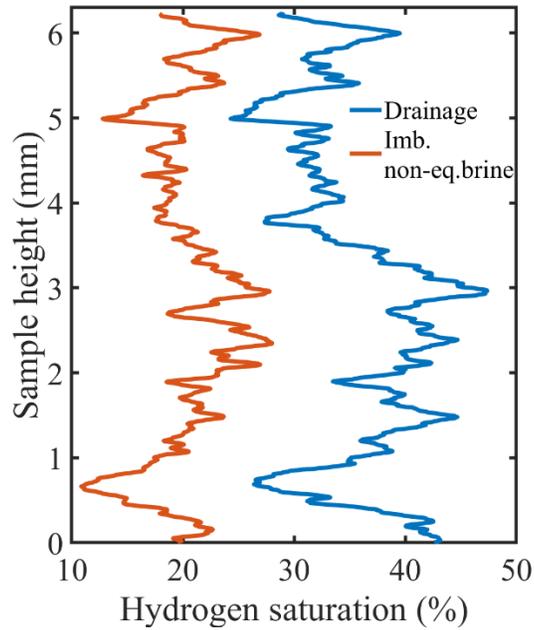

**Figure 8.** Saturation profile along the height of the sample for drainage and imbibition with non-$H_2$-equilibrated brine

## 3.3 A comparison between cycles

Figure 9A shows a comparison of the saturation along the height of the sample for drainage cycles 1 and 2, while Figure 9B shows the pore occupancy. We observe very little difference between the two drainage cycles, indicating a repeatability of the experimental protocol for drainage. The average pore radii and the number of pores occupied by $H_2$ after both drainage cycles (6889 after drainage 1 and 6756 after drainage 2) are similar. The saturation profiles are also similar for both cycles (Figure 9A).

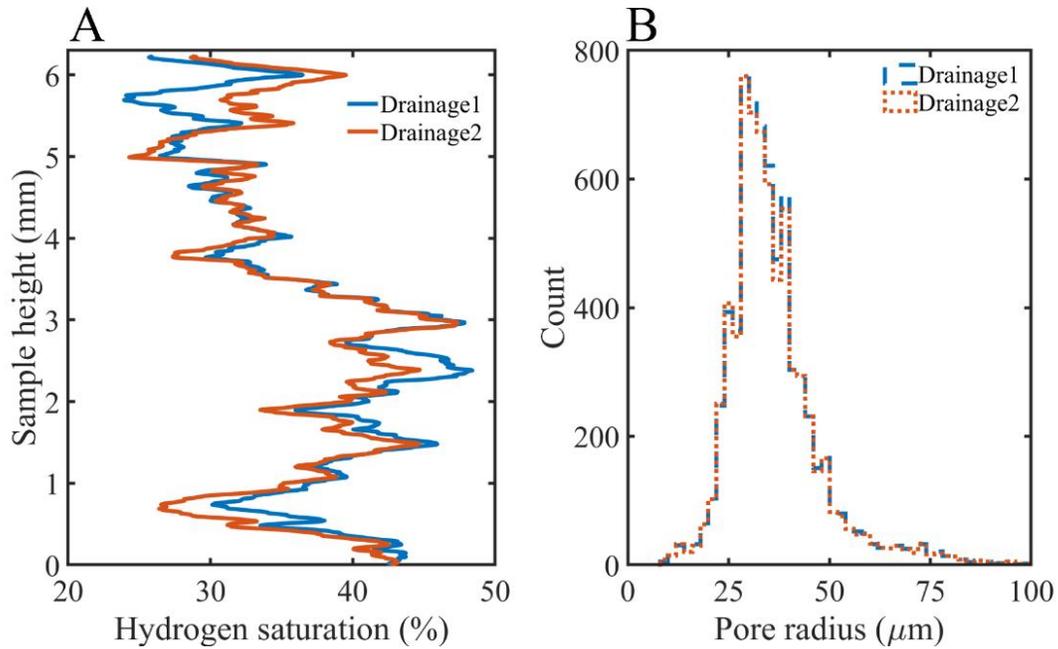

**Figure 9.** $H_2$ saturation profiles (A) and $H_2$ pore occupancy distributions (B) for the two drainage cycles.

However, we observe significant differences between cycle 1 and 2 for imbibition (Figure 10A). The residual $H_2$ saturation in cycle 1 is higher in the lower part of the sample than in cycle 2. This may indicate that some $H_2$ dissolves into the brine during injection at the inlet, lowering the residual saturation in that part of the sample. As the brine travels further into the sample it becomes equilibrated with the $H_2$ and the saturation profile starts to align with the imbibition saturation profile of cycle 1. The pore occupancy profiles (Figure 10B) for both the imbibition processes are similar in terms of the preferred size of pores occupied by $H_2$; however, the number of occupied pores (3331) is significantly lower in the case of imbibition with non-$H_2$-equilibrated brine, as compared to 4333 for imbibition with $H_2$-equilibrated brine, which is expected with a lower residual saturation.

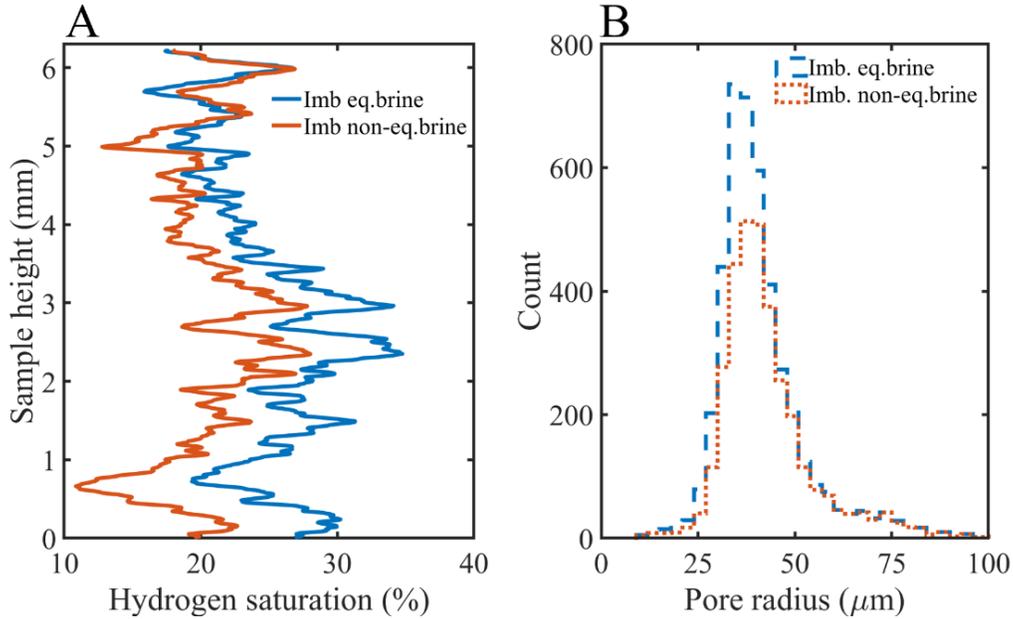

**Figure 10.** $H_2$ saturation profiles (A) and $H_2$ pore occupancy distributions (B) after imbibition with 5 PV of $H_2$-equilibrated brine and non-$H_2$-equilibrated brine.

Image analysis also shows that imbibition with non-$H_2$-equilibrated brine resulted in brine occupying pores which remained filled with $H_2$ in the previous imbibition as shown in Figure 11. Although the solubility of $H_2$ in water is low at the experimental conditions ~ 0.001 g/kg[69], due to the small size of $H_2$ molecules, the mass of $H_2$ occupying the pore space is small. This means that injecting 5 PV of non-$H_2$-equilibrated brine could dissolve up to 7% of the $H_2$ present in the pores. This could explain the lower residual saturation after imbibition with non-$H_2$-equilibrated brine.

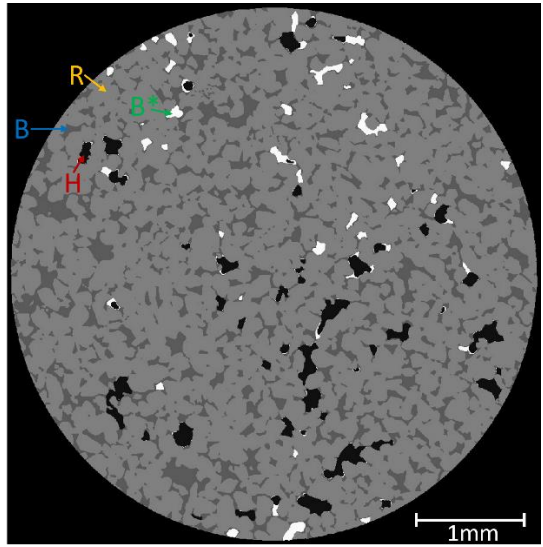

**Figure 11. Snapshot of a slice after imbibition with non-$H_2$-equilibrated brine.** The bright spots (B*) highlight pore spaces that were filled with brine after imbibition with non-$H_2$-equilibrated brine in cycle 2, which were occupied by $H_2$ after imbibition with $H_2$-equilibrated brine in cycle 1.

## 3.4 Pore-scale $H_2$ saturation in context

Quantification of fluid saturations using *in situ* X-ray flow visualization experiments has been frequently reported in the literature for $CO_2$, $N_2$, and oil-brine systems. Some of the earlier reported studies for $CO_2$-brine systems on sandstone samples[36,39,70,71] are plotted in Figure 12 and compared with our results. Three of these earlier studies [36,70,71] are at temperature and pressure conditions similar to those used in this work and the recovery trend is in reasonable agreement with them. Interestingly, the $CO_2$ saturation value closest to our results is for the previous work conducted at ambient conditions[39], as there is a much larger difference between the $\rho/\mu$ value for $H_2$ and $CO_2$ at our experimental conditions ($CO_2$ is a supercritical fluid), compared to $CO_2$ at ambient conditions ($CO_2$ is in the gaseous phase).

The only other reported saturation data for $H_2$-brine system[54] (ambient conditions) is also plotted in Figure 12, where initial $H_2$ saturation of 64% and residual $H_2$ saturation of 41% after 5 PV of brine injection was measured on a Gosford sandstone. Here we observe that the results from both cycle 1 and cycle 2 agree with previously published results in terms of recovery, with cycle 1 slightly above the trend line and cycle 2 slightly below. However, we see significant difference in the $H_2$ saturation values (both initial and residual) compared to earlier reported work for $H_2$[54]. This could be due to the higher temperature and pressure conditions and a larger sub-volume analysed in our work, which provides a more realistic scenario for subsurface storage.

Additionally, the decrease in residual saturation due to dissolution with non-$H_2$-equilibrated brine is likely to be an important parameter for reservoir-scale modelling, especially for UHS reservoirs with cyclic injection as the $H_2$ is stored and then eventually recovered for use in energy production. The initial recovery from the first injection into the reservoir may be low with subsequent cycles producing more $H_2$ as the reservoir brine is slowly equilibrated and loss of $H_2$ gas due to dissolution and trapping decreases. More research is required to investigate this effect and its impact on the fluid saturations.

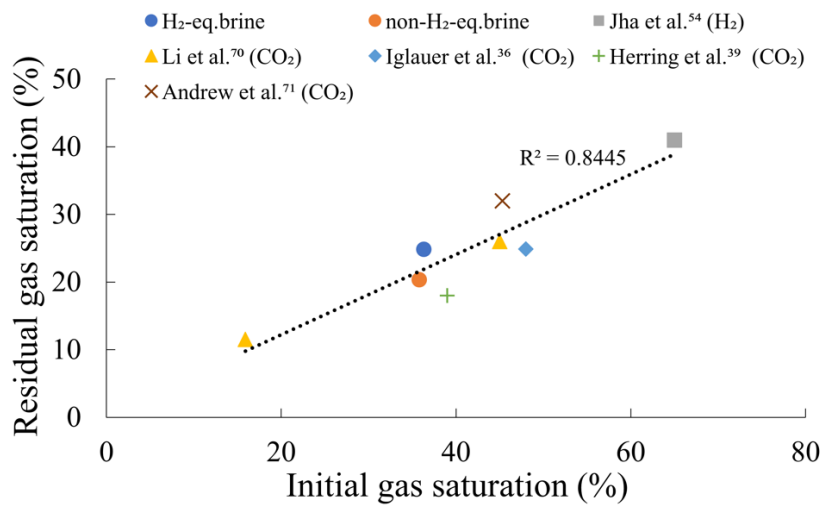

**Figure 12.** Initial vs. residual gas saturation plot for the two imbibition cases in this study compared with literature reported measurements for $H_2$ and $CO_2$ in sandstones from *in situ* X-ray flow visualization experiments.

## 3.5 Contact Angle

Contact angles were determined on a 2.5 mm³ sub-set of each imbibition image. The results (Figure 13) show water-wet conditions with an average contact angle of 53.72° after imbibition with $H_2$-equilibrated brine and 52.72° after imbibition with non-$H_2$-equilibrated brine. These contact angle values agree with the measurements performed using multiple *in situ* methods on a Bentheimer-$H_2$-brine system[32], where average contact angles are reported to be in the range 39.77°-59.75°. The mean contact angle (59.75°), using the same 3D local method[63] used in our study, matches closely with our results. A few other studies[24,47,48] have reported strong water-wet conditions (contact angles between 6° and 43°) regardless of the tested parameters.

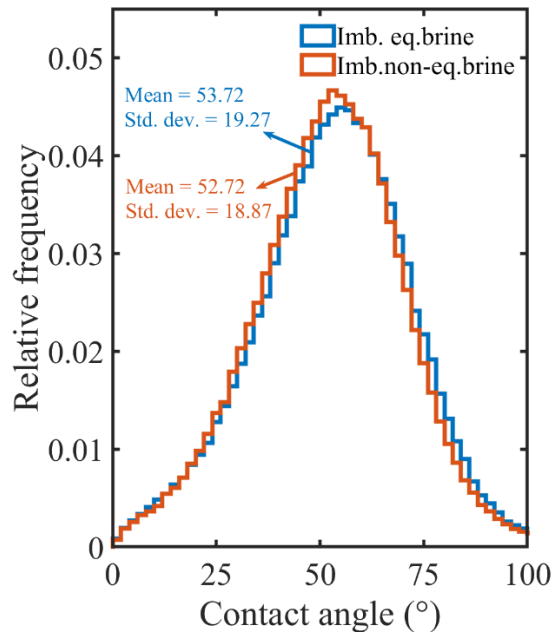

**Figure 13. Contact angle distributions**. After imbibition with $H_2$-equilibrated (cycle 1) and non-$H_2$-equilibrated brine (cycle 2).

## Conclusion

In this study, we have imaged the entrapment of $H_2$ after the injection of both $H_2$-equilibrated and non-$H_2$-equilibrated brine in a real rock at reservoir conditions. Our results can be summarised in three points:

1. $H_2$ is water-wet with an average contact angle of 53.72° and 52.72° obtained after $H_2$-equilibrated and non-$H_2$-equilibrated brine injection respectively, which is in reasonable agreement with published literature[32,34] values. Some other studies [24,47,48] have reported strong water-wet conditions for $H_2$-brine systems. These conditions could enhance snap-off resulting in larger trapping of $H_2$. Determination of the wettability is of prime importance both for trapping and recovery, and more experimental datasets are needed to understand the potential subsurface environment for $H_2$ storage.
2. There was a decrease in residual gas saturation and a decrease in pore occupancy when the sample was imbibed with non-$H_2$-equilibrated brine indicating that dissolution of $H_2$ may be an important consideration for prediction of recovery. Since *in situ* experimental work on $H_2$ at high temperature and pressure conditions is scarce, this effect was not reported in any previous work but can be a critical aspect to consider for future research.
3. The recovery factor trend is consistent with literature values[36,39,70,71] for supercritical $CO_2$ at similar pressure and temperature conditions; however, the saturation values closest to our results are for $CO_2$ at ambient conditions due to relatively smaller $\rho/\mu$ difference.

In summary, even though subsurface systems provide a large volume for $H_2$ to be stored, pore-scale interactions at reservoir conditions need to be further investigated to conclude that these systems are suitable for long-term underground storage. The scope of the research will be expanded to include multiple rock types for our experiments.

## Acknowledgment

We gratefully acknowledge Jim Buckman, Clayton Magill, Paul Miller, Robert Louden, Jim Allison, Emma Samson and Juliane Bischoff for their support in the preparation of the equipment and materials used for our experiments.

## Supplementary Material

### *Segmentation of wet images*

A sample workflow is provided, which was followed to carefully segment each of the scanned image stack. All the image segmentation was performed using Avizo 2021.1 (ThermoFisher Scientific) software. Images below show a small cross-section to clearly show the changes after each step.

State: After imbibition with $H_2$-equilibrated brine

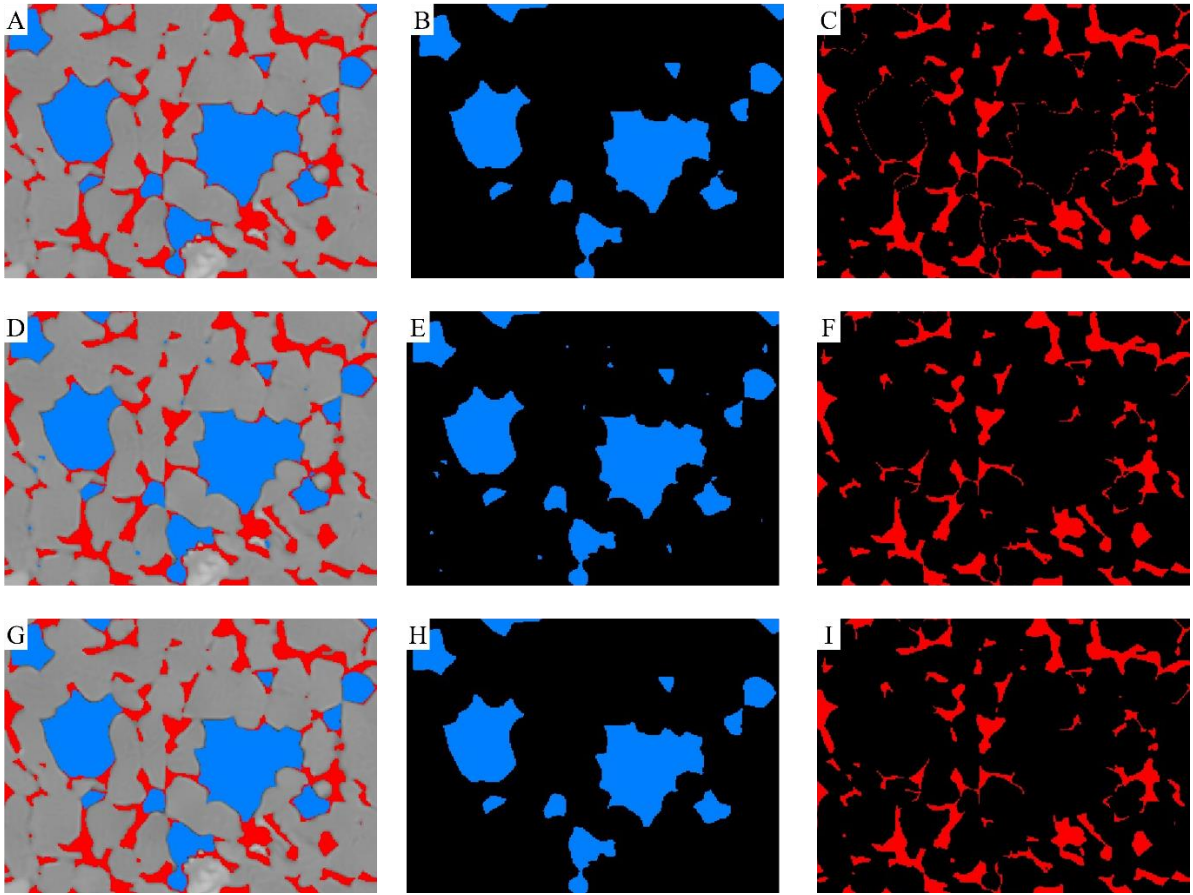

**SM Figure S1. Image segmentation workflow for the wet images.** (**A**) Initial segmentation after watershed with the segmented hydrogen and brine phases overlayed on the rock matrix. (**B**) Initial segmented hydrogen phase obtained using watershed algorithm. (**C**) Brine phase as a result of subtraction of the segmented hydrogen phase from the dry segmented pores of the dry scan. The superficial wetting layers are visible in the image. (**D**) Brine and hydrogen phases overlayed on the rock matrix after removal of small spots from (C). (**E**) Hydrogen phase as a result of assigning the small spots from (C) to the hydrogen phase. This resulted in some small pixels of brine assigned to the hydrogen phase. (**F**) Brine phase after removal of the small spots from (C). (**G**) Final segmented hydrogen and brine phases overlayed on the rock matrix after removal of small spots from (E). (**H**) Final segmented hydrogen phase after removal of small spots from (E). (**I**) Final segmented brine phase obtained by subtracting (H) from the dry segmented pores of the dry scan.